\documentclass{spie}                                                                   
\usepackage{graphicx}
\usepackage{amssymb}

\title{Localization and Pattern Formation in Quantum Physics. \\I.
Phenomena of Localization} 

\author{Antonina N. Fedorova and Michael G. Zeitlin
\skiplinehalf
\supit{}  IPME RAS, St.~Petersburg, V.O. Bolshoj pr., 61, 199178, Russia
}
\authorinfo{http://www.ipme.ru/zeitlin.html,
http://www.ipme.nw.ru/zeitlin.html,\\
E-mail: zeitlin@math.ipme.ru, anton@math.ipme.ru}

\begin{document}

\begin{center}                                                                       
\begin{tabular}{p{160mm}}                                                            
                                                                                     
\begin{center}                                                                       
{\bf\Large                                                                           
LOCALIZATION AND PATTERN FORMATION} \\                                                    
\vspace{5mm}                                                                         
                                                                                     
{\bf\Large IN QUANTUM PHYSICS.}\\                                           
\vspace{5mm}                                                                         
                                                                                     
{\bf\Large I. PHENOMENA OF LOCALIZATION}\\                                                        
                                                                                     
\vspace{1cm}                                                                         
                                                                                     
{\bf\Large Antonina N. Fedorova, Michael G. Zeitlin}\\                               
                                                                                     
\vspace{1cm}

{\bf\large\it                                                                        
IPME RAS, St.~Petersburg,                                                            
V.O. Bolshoj pr., 61, 199178, Russia}\\                                              
{\bf\large\it e-mail: zeitlin@math.ipme.ru}\\                                        
{\bf\large\it e-mail: anton@math.ipme.ru}\\                                          
{\bf\large\it http://www.ipme.ru/zeitlin.html}\\                                     
{\bf\large\it http://www.ipme.nw.ru/zeitlin.html}                                    
                                                                                   
\end{center}                                                                         
                                                                                     
\vspace{1cm}                                                                         
                                                                                     
\begin{abstract}  

In these two related parts we present a set of methods,
analytical
and numerical, which can
illuminate the behaviour of quantum system, especially in the complex
systems, e.g., where the standard "coherent-states" approach cannot be
applied. The key points demonstrating advantages of this approach are:
(i) effects of localization of
possible quantum states, more proper than "gaussian-like states"; 
(ii) effects of non-perturbative multiscales which
cannot be calculated by means of perturbation approaches;
(iii) effects of formation of complex quantum patterns from localized
modes
or classification and possible control of the full zoo of quantum states,
including (meta) stable localized patterns (waveletons).
In this first part we consider the  applications of
numerical-analytical technique based on local nonlinear
harmonic analysis to  quantum/quasiclassical
description of nonlinear (polynomial/rational) dynamical problems
which appear in many areas of physics.
We'll consider calculations of Wigner functions as the
solution of Wigner-Moyal-von Neumann equation(s)
corresponding to polynomial Hamiltonians.
Modeling demonstrates the appearance of (meta) stable patterns generated
by high-localized (coherent) structures or entangled/chaotic behaviour.
We can control the type of behaviour on the level of reduced algebraical
variational system. 
At the end we presented the qualitative definition of the 
Quantum Objects in comparison with their Classical Counterparts, which natural 
domain of definition is the category of multiscale/multiresolution 
decompositions according to the action of internal/hidden symmetry of the proper realization 
of scales of functional spaces (the multiscale decompositions 
of the scales of Hilbert spaces of states). 
It gives rational natural explanation of such pure quantum effects as ``self-interaction''
(self-interference) and instantaneous quantum interaction (transmission of information).

\end{abstract}                                                                       
                                                                                     
\vspace{5mm}                                                                        
                                                                                     
\begin{center}                                                                       
{\large Submitted to Proc. of SPIE Meeting,
 The Nature of Light: What is a Photon?\\
Optics \& Photonics, SP200, }                                                               
{\large San Diego, CA, July-August, 2005}                                        
                                                                                     
\vspace{5mm}

\end{center}                                                                         
\end{tabular}                                                                        
\end{center}                                                                         
\newpage


\maketitle                                  

\begin{abstract}

In these two related parts we present a set of methods,
analytical
and numerical, which can
illuminate the behaviour of quantum system, especially in the complex
systems, e.g., where the standard "coherent-states" approach cannot be
applied. The key points demonstrating advantages of this approach are:
(i) effects of localization of
possible quantum states, more proper than "gaussian-like states"; 
(ii) effects of non-perturbative multiscales which
cannot be calculated by means of perturbation approaches;
(iii) effects of formation of complex quantum patterns from localized
modes
or classification and possible control of the full zoo of quantum states,
including (meta) stable localized patterns (waveletons).
In this first part we consider the  applications of
numerical-analytical technique based on local nonlinear
harmonic analysis to  quantum/quasiclassical
description of nonlinear (polynomial/rational) dynamical problems
which appear in many areas of physics.
We'll consider calculations of Wigner functions as the
solution of Wigner-Moyal-von Neumann equation(s)
corresponding to polynomial Hamiltonians.
Modeling demonstrates the appearance of (meta) stable patterns generated
by high-localized (coherent) structures or entangled/chaotic behaviour.
We can control the type of behaviour on the level of reduced algebraical
variational system. 
At the end we presented the qualitative definition of the 
Quantum Objects in comparison with their Classical Counterparts, which natural 
domain of definition is the category of multiscale/multiresolution 
decompositions according to the action of internal/hidden symmetry of the proper realization 
of scales of functional spaces (the multiscale decompositions 
of the scales of Hilbert spaces of states). 
It gives rational natural explanation of such pure quantum effects as ``self-interaction''
(self-interference) and instantaneous quantum interaction (transmission of information).

\end{abstract}

\vspace{5mm}

\section{INTRODUCTION}

It is obviously that the current (and the nearest future) level of possible 
experiments in the area of quantum physics as a whole
and quantum computers as some quintessence and important region of applications,
as well as the level of the phenomenological modeling overtake the present level of 
mathematical/theoretical description [1].
Considering, for example, the problem of description of possible/realizable states,
one hardly believes that plane waves or/and (squeezed) gaussians/coherent states 
exhausted all needful things, demanded, e.g.,
for full description/construction of the possible quantum CPU-like devices.
Complexity of the proper zoo of states, including entangled/chaotic states, 
is still far from clear understanding which can provide practical controllable realization.
It seems that the first thing, one needs to do in this direction, is related to 
a possible reasonable extension of understanding of the quantum dynamics in a whole volume.
One needs to sketch up the underlying ingredients of the theory (spaces of  states, observables, measures,
classes of smothness, quantization set-up etc) in an attempt to provide the maximally extendable
but at the same time really calculable and realizable description of the dynamics of quantum world.
Generic idea is very trivial: it is well known that the idea of ``symmetry'' is the key ingredient of any
reasonable physical theory from classical (in)finite dimensional (integrable) Hamiltonian 
dynamics to different sub-planckian models based on strings/branes/orbifolds etc.
During century kinematical, dynamical and hidden symmetries
played the key role in our understanding of physical process.
Roughly speaking, the representation theory of underlying 
symmetry (classical or quantum, groups or (bi)algebras, finite or infinite dimensional, continuous 
or discrete) is a proper instrument for description of (orbital) dynamics.
Starting point for us is a possible model for (continuous) ``qubit'' with subsequent 
description of the whole zoo of possible realizable/controllable states/patterns 
which may be useful from the point of view of quantum experimentalists and/or engineers.
The proper representation theory is well known as ``local nonlinear harmonic analysis'',
in particular case of simple underlying symmetry--affine group--aka wavelet analysis.
From our point of view the advantages of such approach are as follows:

i) natural realization of localized states in any proper functional relalization of 
(Hilbert) space of states.

ii) hidden symmetry of choosen realization of proper functional model provides 
the (whole) spectrum of possible states via the so-called multiresolution decomposition.

So, indeed, the hidden symmetry (non-abelian affine group in the simplest case) 
of the space of states via proper representation theory generates the physical spectrum
and this procedure depends on the choice of the functional realization of the space of states.  
It explicitly demonstrates that the structure and properties of the functional realization of the
space of states are the natural properties of physical world at the same level of importance as
a particular choice of Hamiltonian, or equation of motion, or action principle (variational method).      

At the next step we need to consider the consequences of 
our choice i)-ii) for the algebra of observables.
In this direction one needs to mention the class of operators we are interested in to
present proper description for a class of maximally generalized but reasonable class of problems.
It seems that it must be the pseudodifferential operators, especially 
if we underline that in the spirit of points i)-ii) above we need to take Wigner-Weyl framework 
as generic for constructing basic quantum equations of motions.
It is obviously, that consideration of symbols of operators insted of operators themselves
is the starting point as for the mathematical theory of pseudodifferential operators as for 
quantum dynamics formulated in the language of Wigner-like equations.
It should be noted that in such picture we can naturally include the effects of self-interaction on the way of 
construction and subsequent analysis of nonlinear quantum models.
So, our consideration will be in the framework of (Nonlinear) Pseudodifferential Dynamics
($\Psi DOD$).
As a result of i)--ii), we'll have:

iii) most sparse, almost diagonal, representation for wide 
class of operators included in the set-up of the whole problems.

It's possible by using the so-called Fast Wavelet Transform representation for algebra of observables..

Then points i)--iii) provide us by

iv) natural (non-perturbative) multiscale decomposition 
for all dynamical quantities, as states as observables. 

The simplest and proper case we'll have in Wigner-Weyl representation.

Existence of such internal multiscales with different dynamics at each scale and transitions, interactions,
and intermittency between scales demonstrates that quantum mechanics, nevertheless  its
linear structure, is really a serious part of physics. 
It seems, that this underlying quantum complexity is a result of transition 
by means of (still unclear) procedure of quantization
from complexity related to nonlinearity of classical counterpart to the rich 
pseudodifferential (more exactly, microlocal) structure on the quantum side.    
For the reasons, explained later, we prefer to divide all possible configurations
related to possible solutions of our quantum equation of motion (Wigner-like equations, mostly)
into two classes: 

(a) standard solutions
(b) controllable solutions.

The meaning of the latter will be explained  in part II, but,
anyway, the whole zoo of solutions consists of possible patterns, including
very important ones from the point of view of underlying physics:

v) localized modes (basis modes, eigenmodes) and constructed from them chaotic or/and entangled,
decoherent (if we change Wigner equation for (master) Lindblad one) patterns.

It should be noted that these bases modes are (non) linear in contrast with usual ones
because they come from (non) abelian basis group while the usual Fourier (commutative) 
analysis starts from $U(1)$ abelian modes (plane waves).
They are really ``eigen'' but in sense of decomposition of representation of the underlying
hidden symmetry group which generates the multiresolution decomposition.
Zoo of patterns is built from these modes by means of variational procedures (there is a lot of)
more or less standard in mathematical physics.
It allows to control the convergence from one side but, what is more important,

vi) to consider the problem of the control of patterns (types of behaviour) on the level
of reduced (variational) algebraical equations.

We need to mention that it's possible to change the simplest generic group of hidden internal symmetry 
from the affine (translation and dilations) to much more general, 
but, in any case, this generic 
symmetry will produce the proper 
natural high localized eigenmodes, as well as the decomposition of the functional realization
of space of states into the proper orbits; and all that allows to compute dynamical consequence 
of this procedure, i.e. pattern formation, and, as a result, to classify the whole spectrum of proper states.
For practical reasons controllable patterns (with prescribed behaviour) are the most useful.
We mention the so-called waveleton-like pattern, as the most important one.
We use the following allusion in the space of words:

\{waveleton\}:=\{soliton\}  $\bigsqcup$   \{wavelet\}

It means:

vii) waveleton $\approx$ (meta)stable localized (controllable) pattern

To summarize, we may say that approach described below allows us 

viii) to solve wide classes of general 
$\Psi DOD$ problems, including generic for quantum physics Wigner-like equations, and

ix) to present the analytical/numerical realization for physically interesting patterns.

It should be noted  the effectiveness of numerical realization of this program
(minimal complexity of calculations) as additional advantage.
So, items i)-ix) point out all main features of our approach [2]-[8].

\section{MOTIVES}
\subsection{Class of Models}

Here we describe a class of problems which can be analysed by methods described in Introduction.
We start from individual dynamics and finish by (non)-equilibrium ensembles.
All models are belonged to $\Psi DOD$ class and described by finite or infinite 
(named hierarchies in such cases) system of 
$\Psi DOD$ equations.

\begin{itemize}
\item[a).] Individual classical/quantum mechanics ($cM/qM$) (linear/nonlinear; $\{cM\}\subset\{qM\}$), 
\\

($\ast$ - Quantization of) Polynomial Hamiltonians:

\begin{equation}
H(p,q,t)=\sum_{i.j}a_{ij}(t)p^iq^j
\end{equation}

\item[b).] QFT-like models in framework of the second quantization (dynamics in Fock spaces).
\item[c).] Classical (non) equilibrium ensembles via BBGKY Hierarchy. 
(with reductions to different forms of Vlasov-Maxwell/Poisson equations). 
\item[d).] Wignerization of a): Wigner-Moyal-Weyl-von Neumann-Lindblad.
\item[e).] Wignerization of c): Quantum (Non) Equilibrium Ensembles.
\end{itemize}

Important remark, points a)-e), are considered in $\Psi$DO 
picture of  (Non)Linear $\Psi$DO Dynamics (surely, all $qM\subset\Psi DOD$).
Dynamical variables/observables are the symbols of operators or functions; in case
of ensembles, the main set of dynamical variables consists 
of partitions (n-particle partition functions).

\subsection{Effects (what we are interested in)}

\begin{itemize}
\item[i).] Hierarchy of internal/hidden scales (time, space, phase space).
\item [ii).] Non-perturbative multiscales: 
from slow to fast contributions,
from the coarser to the finer level of resolution/de\-composition.
\item [iii).] Coexistence of hierarchy of multiscale dynamics with transitions between scales.
\item [iv).] Realization of the key features of the complex quantum 
world such as the existence of chaotic and/or entangled 
states with possible destruction in ``open/dissipative'' regimes due to interactions with
quantum/classical environment and transition to decoherent states.

\end{itemize}

At this level we may interpret the effect of misterious entanglement or ``quantum interaction''  
as a result of simple interscale interaction or intermittency (with allusion to hydrodynamics)
as the mixing of orbits generated by multiresolution representation of hidden underlying symmetry.
Surely, the concrete realization of such a symmetry is a natural physical property 
of the physical model as well as the space of representation and its proper functional 
realization.
So, instantaneous interactions (or transmission of ``quantum bits'' or ``teleportation'') 
materialize not in the physical space-time variety but in the space of representation of hidden symmetry
along the orbits/scales constructed by proper representations. 
Dynamical/kinematical principles of usual space-time varieties, definitely, don't cover
kinematics of internal quantum space of state or, in more weak formulation, 
we still havn't such explicit relations.

One additional important comment:
as usually in modern physics, we have the hierarchy of underlying symmetries; so
our internal symmetry on functional realization of space 
of states is really not more than kinematical, because  much more rich
algebraic structure, related to operator Cuntz algebra and quantum groups, is hidden inside.
The proper representations can generate much more interesting effects than ones described above.
We'll consider it elsewhere but mention here only how it can be realized by the existing  
functorial maps between proper categories:

\{QMF\} $\longrightarrow$ Loop groups $\longrightarrow$ Cuntz operator algebra 
$\longrightarrow$ Quantum Group structure,
where \{QMF\} are the so-called quadratic mirror filters generating the realization of 
multiresolution decomposition/representation in any functional space; loop group
is well known in many areas of physics, e.g. soliton theory, strings etc, 
roughly speaking, its algebra coincides with Virasoro algebra; Cuntz operator algebra
is universal $C^*$ algebra generated by N elements with two relations between them;
Quantum group structure (bialgebra, Hopf algebra, etc) is well known in many areas 
because of its universality.
It should be noted the appearance of natural Fock structure inside this functorial sequence
above with the creation operator realized as some generalization of Cuntz-Toeplitz isometries.
Surely, all that can open a new vision of old problems and bring new possibilities. 

We finish this part by the following qualitative definitions of objects, 
which description and understanding 
in different physical models is our main goal in these two papers. 

\begin{itemize}
\item
By a localized state (localized mode) 
we mean the corresponding (particular) solution (or generating mode) which 
is localized in maximally small region of the phase (as in c- as in q-case) space.

\item
By an entangled/chaotic pattern we mean some solution (or asymptotics of solution) 
which has random-like distributed energy (or information) spectrum in a full domain of definition. 
In quantum case we need to consider additional entangled-like patterns, roughly speaking,
which cannot be separated into pieces of sub-systems.

\item
By a localized pattern (waveleton) 
we mean (asymptotically) (meta) stable solution localized in 
relatively small region of the whole phase space (or a domain of definition). 
In this case the energy is distributed during some time (sufficiently large) 
between only a few  localized modes (from point 1). 
We believe, it is a good image for plasma in a fusion state (energy confinement)
or model for quantum continuous ``qubit'' or a result of the process of decoherence 
in open quantum system when the full entangled state degenerates 
into localized (quasiclassical) pattern.

\end{itemize}

\subsection{Methods}

\begin{itemize}
\item[i).] Representation theory of internal/hidden/underlying symmetry,
Kinematical, Dynamical, Hidden.

\item[ii).] Arena (space of representation): proper functional realization of (Hilbert) space of states.

\item[iii).]
Harmonic analysis on (non)abelian group of internal symmetry.
Local/Nonlinear (non-abelian) Harmonic Analysis 
(e.g, wavelet/gabor etc. analysis) instead of linear non-localized $U(1)$ Fourier analysis.
Multiresolution (multiscale) representation.
Dynamics on proper orbit/scale (inside the whole hierarchy of multiscales) in functional space.
The key ingredients are 
the appearance of multiscales (orbits) and the existence of high-localized 
natural (eigen)modes [10].

\item[iv).]
Variational formulation (control of convergence, reductions 
to algebraic systems, control of type of behaviour).

\end{itemize}

\section {SET-UP/FORMULATION} 

Let us consider the following $\Psi$DOD dynamical problem

\begin{equation}
L^j\{Op^i\}\Psi=0,
\end{equation}
described by a finite or infinite number of 
equations which include general classes of operators
$Op^i$ such as differential, integral, pseudodifferential etc

Surely, all Wigner-like equations/hierarchies are inside.

The main objects are:

\begin{itemize}
\item[i).] (Hilbert) space of states, $H=\{\Psi\}$, with a proper functional 
realization, e.g.,: $L^2$, Sobolev, Schwartz,
$C^0$, $C^k$, ... $C^\infty$, ...; \\
Definitely, $L^2(R^2)$, $L^2(S^2)$, $L^2(S^1\times S^1)$, $L^2(S^1\times S^1\ltimes Z_n)$
are different objects.
     
\item[ii).]
Class of smoothness. The proper choice determines natural consideration of dynamics\\ 
with/without Chaos/Fractality property.

\item[iii).] Decompositions 

\begin{equation}
\Psi\approx\sum_ia_ie^i
\end{equation} 

via high-localized bases (wavelet families, generic wavelet packets etc), 
frames, atomic decomposition
(Fig. ~1) with the following 
main properties:
(exp) control of convergence, maximal rate of convergence  for any $\Psi$ in any $H$.

\begin{figure}
\begin{center}
\begin{tabular}{c}
\includegraphics*[width=100mm]{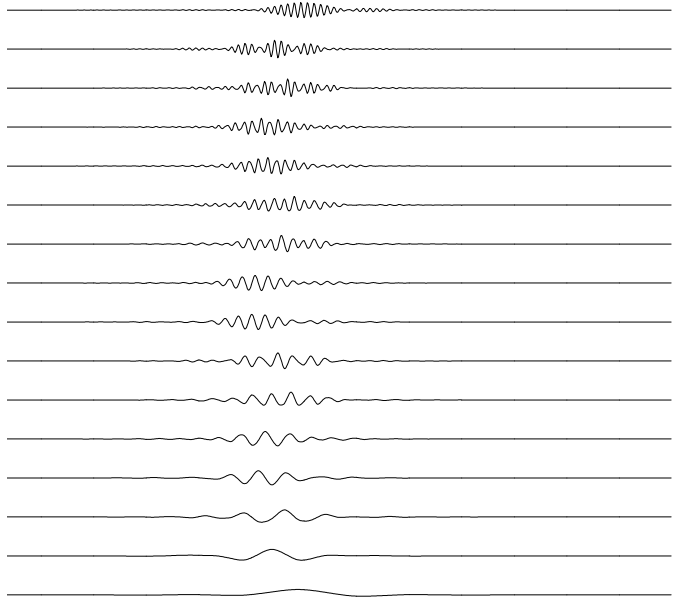}
\end{tabular}
\end{center}
\caption{Localized modes.}
\end{figure}

\item[iv).] Observables/Operators (ODO, PDO, $\Psi$DO, SIO,...,
Microlocal analysis of Kashiwara-Shapira (with change from functions to sheafs))
satisfy the main property -- the matrix representation in localized bases
\begin{equation}
<\Psi|Op^i|\Psi>
\end{equation}

is maximum sparse: 

\begin{displaymath}
\left(\begin{array}{cccc}
D_{11} & 0 &0 & \ldots\\
0    & D_{22} & 0 & \ldots\\
0    & 0    & D_{33} & \ldots\\
\vdots & \vdots & \vdots & \ddots
\end{array} \right)
\end{displaymath}

This almost diagonal structure is provided by the so-called Fast Wavelet Transform.

\item[v).] Measures: multifractal wavelet measures $\{\mu_i\}$ together with the class of smothness
are very important for analysis of complicated analytical behaviour.

\item[vi).] Variational/Pojection methods, from Galerkin to 
Rabinowitz minimax, Floer (in symplectic case of Arnold-Weinstein curves with
preservation of Poisson/symplectic structures).
Main advantages are the reduction to algebraic systems, which provides a tool for the smart   
subsequent control of behaviour and control of convergence.

\item[vii).] Multiresolution or multiscale decomposition, $MRA$ (or wavelet microscope) 
consists of the understanding and choosing of
   \begin{itemize}
   \item[1).] (internal) symmetry structure, e.g., 

affine group = \{translations, dilations\} or 
many others... and construction of
   \item[2).] representation/action of this symmetry on $H=\{\Psi\}$.
    
As a result of such hidden coherence together with using point v) we'll have:
     \begin{itemize}
     \item[a).] LOCALIZED BASES  $\qquad$ b). EXACT MULTISCALE DECOMPOSITION
with the best convergence properties and real evaluation of the rate of convergence
via proper ``multi-norms''
      \end{itemize}
      \end{itemize}
Figures~2, 3, 5, 6 demonstrate MRA decompositions for one- and multi-kicks while Figures
4 and 7 present the same for the case of the generic simple fractal model, Riemann--Weierstrass
function.

\item[viii).] Effectiveness of proper numerics: CPU-time, HDD-space, minimal complexity of
algorithms, and (Shannon) entropy of calculations, are provided by points i)-vii) above.

\item[ix).] Quantization via $\ast$ star product or Deformation Quantization.
We'll use it on Wigner-Weyl-Moyal naive level.    
\end{itemize}

Finally, such Variational-Multiscale approach based on points i)-viii) 
provides us by the full ZOO of PATTERNS: 
LOCALIZED, CHAOTIC/ENTANGLED, etc. In next Section and in the Part II we'll consider 
details for important cases of Wigner-like equations.

\section{PATTERN FORMATION IN WIGNER-MOYAL DYNAMICS}

So, in this Section shortly and in the Part II [9] in details, we'll 
consider the  applications of  
nu\-me\-ri\-cal\--\-ana\-ly\-ti\-cal technique based on local nonlinear harmonic analysis
(wavelet analysis [10]) to  (quasiclassical) 
quantization of nonlinear (polynomial/rational) dynamical problems 
which appear in many areas of physics [1].
Our starting point is the general point of view of deformation 
quantization approach at least on
naive Moyal level [1].
So, 
for quantum calculations we need to find
an associative (but non-commutative) star product $*$ on the space of formal power series in $\hbar$ with
coefficients in the space of smooth functions such that
\begin{equation}
f * g =fg+\hbar\{f,g\}+\sum_{n\ge 2}\hbar^n B_n(f,g). 
\end{equation}
In this paper we consider calculations of Wigner functions $W(p,q,t)$ (WF) as the solution
of some sort of Wig\-ner equations [1]:
\begin{eqnarray}
i\hbar\frac{\partial}{\partial t}W = H * W - W * H
\end{eqnarray}
corresponding to polynomial Hamiltonians  $H(p,q,t)$.
In general, equation (6)
is nonlocal (pseudodifferential) for arbitrary Hamiltonians but in case of 
polynomial Ha\-mil\-to\-ni\-ans
we have only a finite number of terms in the corresponding series.
For example, in stationary case, after Weyl-Wigner mapping we have the following 
equation on WF in c-numbers [1]:
\begin{eqnarray}
\Big( \frac{p^2}{2m}+\frac{\hbar}{2i}\frac{p}{m}\frac{\partial}{\partial q}-
 \frac{\hbar^2}{8m}\frac{\partial^2}{\partial q^2}\Big)W(p,q)+
 U\Big(q-\frac{\hbar}{2i}\frac{\partial}{\partial p}\Big)W(p,q)=\epsilon W(p,q)
\end{eqnarray}
and after expanding the potential $U$ into the Taylor 
series we have two real finite partial differential equations.
Our approach is based on extension of our variational-wavelet approach [2]-[8].
Wavelet analysis is some set of mathematical methods, which gives us the possibility to
work with well-localized bases in functional spaces and gives maximum sparse
forms for the general type of operators (differential, integral, pseudodifferential) 
in such bases.
We decompose the solutions of (6), (7) according to underlying hidden scales as 
\begin{eqnarray}
W(t,q,p)=\displaystyle\bigoplus^\infty_{i=i_c}\delta^iW(t,q,p)
\end{eqnarray}
where value $i_c$ corresponds to the coarsest level of resolution
$c$ in the full multiresolution decomposition
\begin{equation}
V_c\subset V_{c+1}\subset V_{c+2}\subset\dots.
\end{equation}
As a result, the solution of any Wigner-like equations has the 
following mul\-ti\-sca\-le\-/mul\-ti\-re\-so\-lu\-ti\-on decomposition via 
nonlinear high\--lo\-ca\-li\-zed eigenmodes, 
which corresponds to the full multiresolution expansion in all underlying  
scales starting from coarsest one
(polynomial tensor bases, variational machinery 
etc will be considered in details in Part II [9]; ${\bf x}=(x,y,p_x,p_y,...)$):
\begin{eqnarray}\label{eq:z}
&&W(t,{\bf x})=\sum_{(i,j)\in Z^2}a_{ij}{\bf U}^i\otimes V^j(t,{\bf x}),\\
&&V^j(t)=V_N^{j,slow}(t)+\sum_{l\geq N}V^j_l(\omega_lt), \quad \omega_l\sim 2^l \nonumber\\
&&{\bf U}^i({\bf x})={\bf U}_M^{i,slow}({\bf x})+
\sum_{m\geq M}{\bf U}^i_m(k_m{\bf x}), \ k_m\sim 2^m\nonumber
\end{eqnarray}

Representation (10) gives the expansion into the slow part
and fast oscillating parts for arbitrary N, M.  So, we may move
from coarse scales of resolution to the 
finest one, along the orbits generated by actions of hidden internal symmetry group,
to obtain more detailed information about our dynamical process.
In contrast with different approaches representation (10) doesn't use perturbation
technique or linearization procedures.
So, by using wavelet bases with their best (phase) space/time      
localization properties we can describe complex structures in      
quantum systems with complicated behaviour.
Modeling demonstrates the formation of different patterns, 
including chaotic/entangled from
high-localized coherent structures.
Our (nonlinear) eigenmodes are more realistic for the modeling of 
nonlinear classical/quantum dynamical process  than the corresponding linear gaussian-like
coherent states. Here we mention only the best convergence properties of expansions 
based on wavelet packets, which  realize the minimal Shannon entropy property
and exponential control of convergence based on the norm introduced in [7], [8].
Fig.~9 and Fig.~10, ~11 show the contributions to WF from 
localized eigenmodes in dimensions one and two, respectively,
while Fig. 11--14  gives 
the representations for full solutions, constructed
from the first 6 scales via MRA and demonstrate stable localized 
pattern formation (waveleton) and chaotic/entangled behaviour. 
For comparison, Fig.~8, demonstrates direct modeling [10] of WF for the case of three localized
modes which are belonged to generic family of wavelet packets. 

Let us finish with some phenomenological description which can be 
considered as an introduction to Part 2 [9] and at the same time as 
an attempt of qualitative description of the quantum dynamics 
as a whole and in comparison with its classical counterpart.
It's possible to take for reminiscence the famous Dirac's phrase that ``electron can interact only 
itself via the process of quantum interference''.

Let $G$ be the hidden/internal symmetry group on the spaces of quantum states which generates via MRA
(9) the multiscale/multiresolution representation for all 
dynamical quantities, unified in object $O(t)$, such as states, observables, partitions: 
$O^i(t)=\{\psi^i(t), Op^i(t), W_n^i(t)\}$, where $i$ is the proper scale index.
Then, the following (commutative) diagram represents the details of quantum life from 
the point of view of representations of $G$ on the choosen functional realization which leads to
decomposition of the whole quantum evolution into the proper orbits or scales corresponding 
to the proper level of resolution. Morphisms $W(t)$ describe Wigner-Weyl evolution in the 
algebra of symbols, while the processes of interactions with open World, such as
the measurement or decoherence, correspond to morphisms (or even functors)
which transform the infinite set of scales
characterizing the quantum object into finite ones, sometimes consisting of one element
(demolition/destructive measurement).

\begin{eqnarray}
&&\qquad\qquad\qquad W(t) \nonumber\\
&&\lbrace O^i(t_1) \rbrace \qquad  \longrightarrow   \qquad \lbrace O^j(t_2) \rbrace \nonumber\\
&&\qquad\qquad\nonumber\\
&&\downarrow m(t_1) \qquad \qquad \qquad  \downarrow m(t_2) \nonumber\\
&&\qquad\qquad\qquad \widetilde{W(t)} \nonumber\\
&&\lbrace O^{i_c}(t_1) \rbrace \qquad  \longrightarrow  \qquad \lbrace O^{j_c}(t_2) \rbrace\nonumber
\end{eqnarray}
where reduced morphisms $\widetilde{W(t)}$ correspond to (semi)classical or 
quasiclassical evolution. 
 
So, qualitatively,

{\bf Quantum Objects} can be represented by an infinite or enough large set of coexisting and
interacting subsets like (8), (9) (Figures 1-7 for some allusion).

while 

{\bf Classical Objects} can be described by one or few only levels of resolution
with (almost) supressed interscale self-interaction.

It's possible to consider Wigner functions as some measure of the quantum character of the system:
as soon as it will be positive, we arrive to classical regime and 
no more reasons to have the full hierarchy decomposition in the representations like (8).
So, Dirac's self-interference is nothing but the multiscale mixture/intermittency.
Definitely, the degree of this self-interaction leads to different qualitative types 
of behaviour, such as localized quasiclassical states, separable, entangled, chaotic etc.
At the same time the instantaneous quantum interaction or transmission of information from Alice 
to Bob takes place not in the physical kinematical space-time but in Hilbert spaces of states
in their proper functional realization where there is a different kinematic life.
To describe a set of Quantum Objects we need to realize our Space of States (Hilbert space)
not as one functional space but as the so-called and well known in 
mathematics scale of spaces, e.g. $B^s_{p,q}$, $F^s_{p,q}$ [11].
The proper multiscale decomposition for the scale of space provides us by the 
method of description of the set of quantum objects in case if the ``size'' of 
one Hilbert space is not enough to describe the complicated internal World.
We'll consider it elsewhere, while in the second part [9] we consider the one-scale case
(to avoid possible misunderstanding we need take into account that one-scale case 
is also described by an infinite scale of spaces (9), but it's internal 
decomposition of the unique, attached to the problem, Hilbert space).  

\vspace{5mm}

\acknowledgments

We are very grateful to Prof. Chandrasekhar Roychoudhuri, University of Connecticut,
Ms. Jenny Woods, Mrs. Kristi Kelso, SPIE (International Society for Optical Engineering)
for kind attention and help, and
United States Air Force Office for Scientific Research, and Nippon Sheet Glasses Co. Ltd., 
and SPIE for financial support provided our participation in SPIE Meeting, 
San Diego, July-August, 2005.

\newpage

\begin{twocolumn}

\begin{figure}
\begin{center}
\begin{tabular}{c}
\includegraphics[width=60mm]{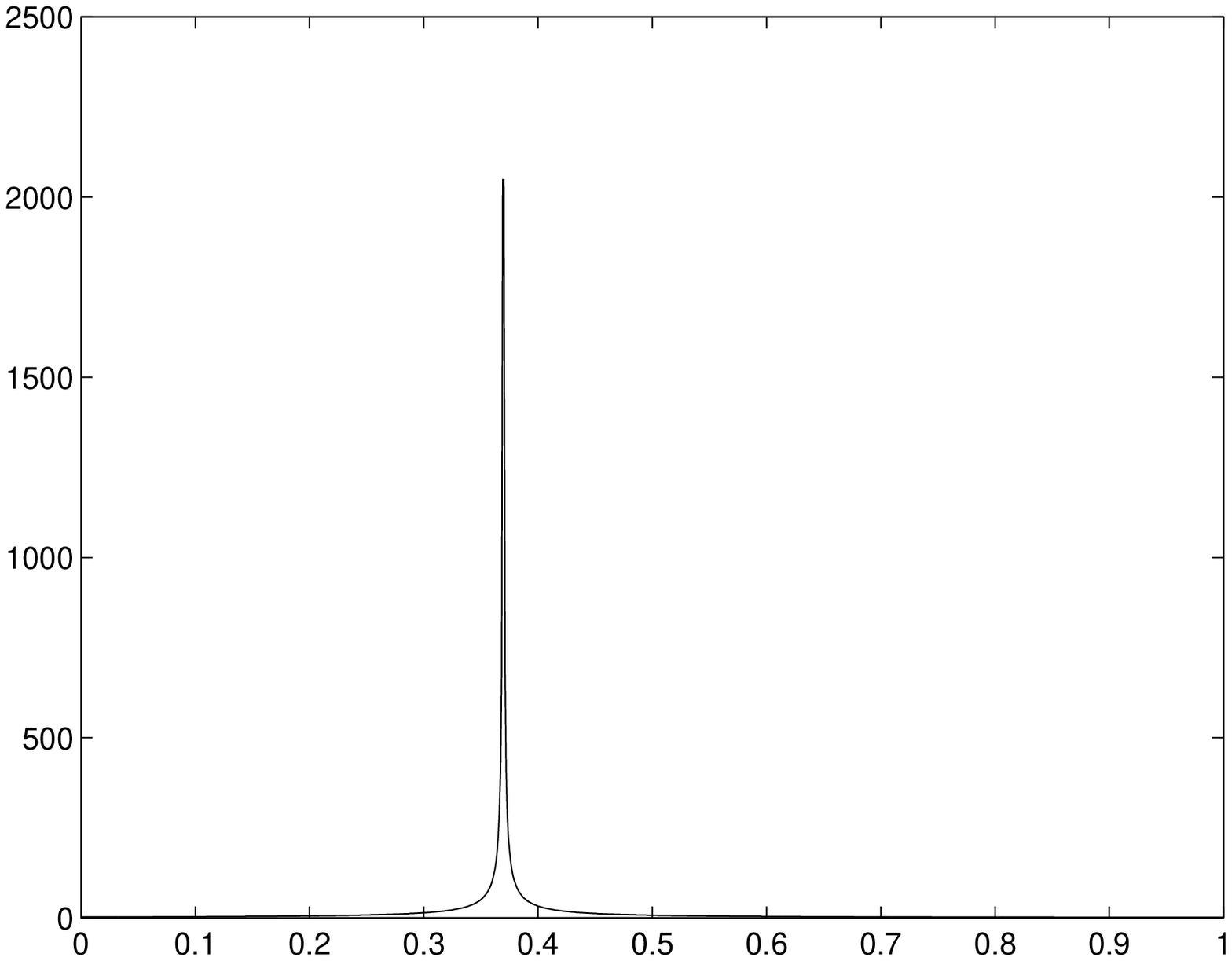}
\end{tabular}
\end{center}
\caption{Kick.}
\end{figure}

\begin{figure}
\begin{center}
\begin{tabular}{c}
\includegraphics[width=60mm]{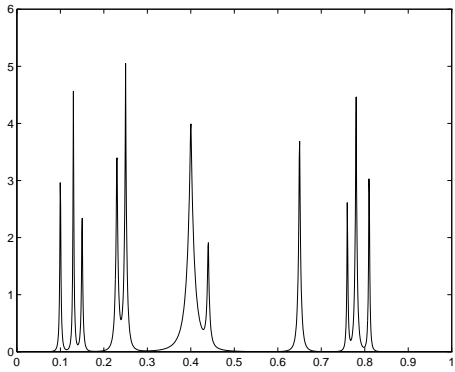}
\end{tabular}
\end{center}
\caption{Multi-Kicks.}
\end{figure}

\begin{figure}
\begin{center}
\begin{tabular}{c}
\includegraphics[width=60mm]{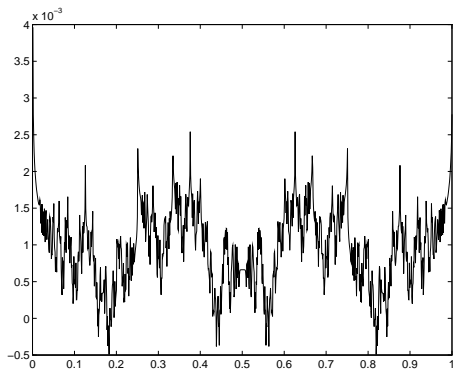}
\end{tabular}
\end{center}
\caption{RW-fractal.}
\end{figure}

\begin{figure}
\begin{center}
\begin{tabular}{c}
\includegraphics[width=60mm]{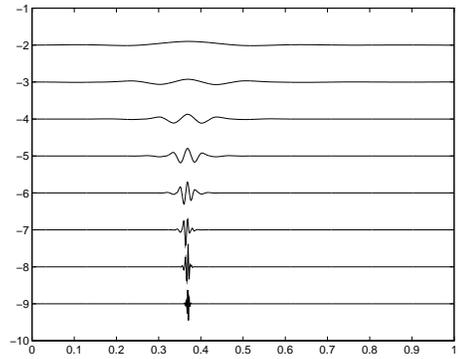}
\end{tabular}
\end{center}
\caption{MRA for Kick.}
\end{figure}

\begin{figure}
\begin{center}
\begin{tabular}{c}
\includegraphics[width=60mm]{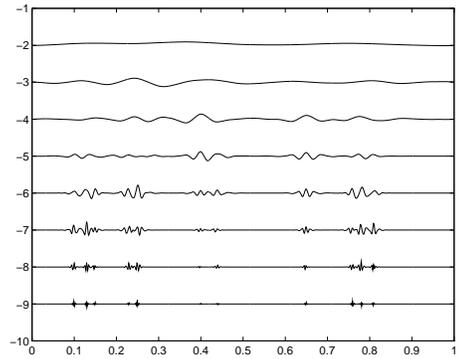}
\end{tabular}
\end{center}
\caption{MRA for Multi-Kicks.}
\end{figure}

\begin{figure}
\begin{center}
\begin{tabular}{c}
\includegraphics[width=60mm]{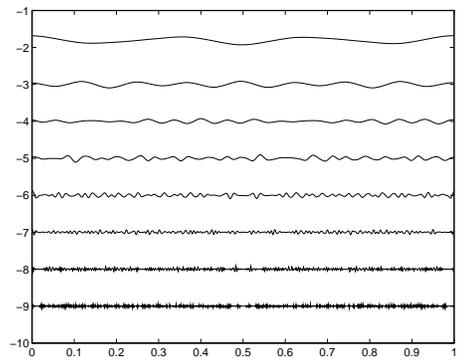}
\end{tabular}
\end{center}
\caption{MRA for RW-fractal.}
\end{figure}

\end{twocolumn}

\onecolumn

\begin{figure}
\begin{center}
\begin{tabular}{c}
\includegraphics[width=100mm]{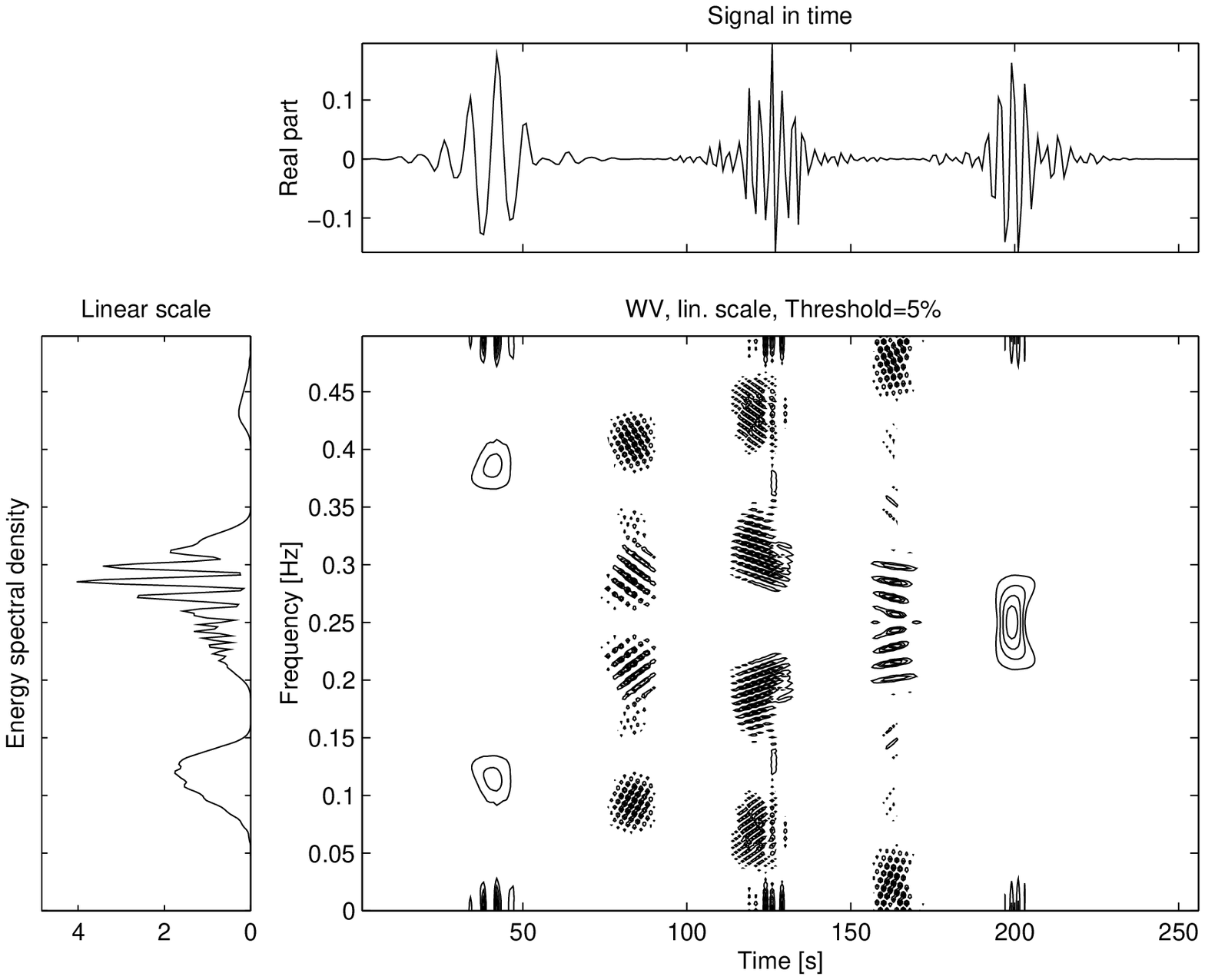}
\end{tabular}
\end{center}
\end{figure}

\begin{figure}
\begin{center}
\begin{tabular}{c}
\includegraphics[width=100mm]{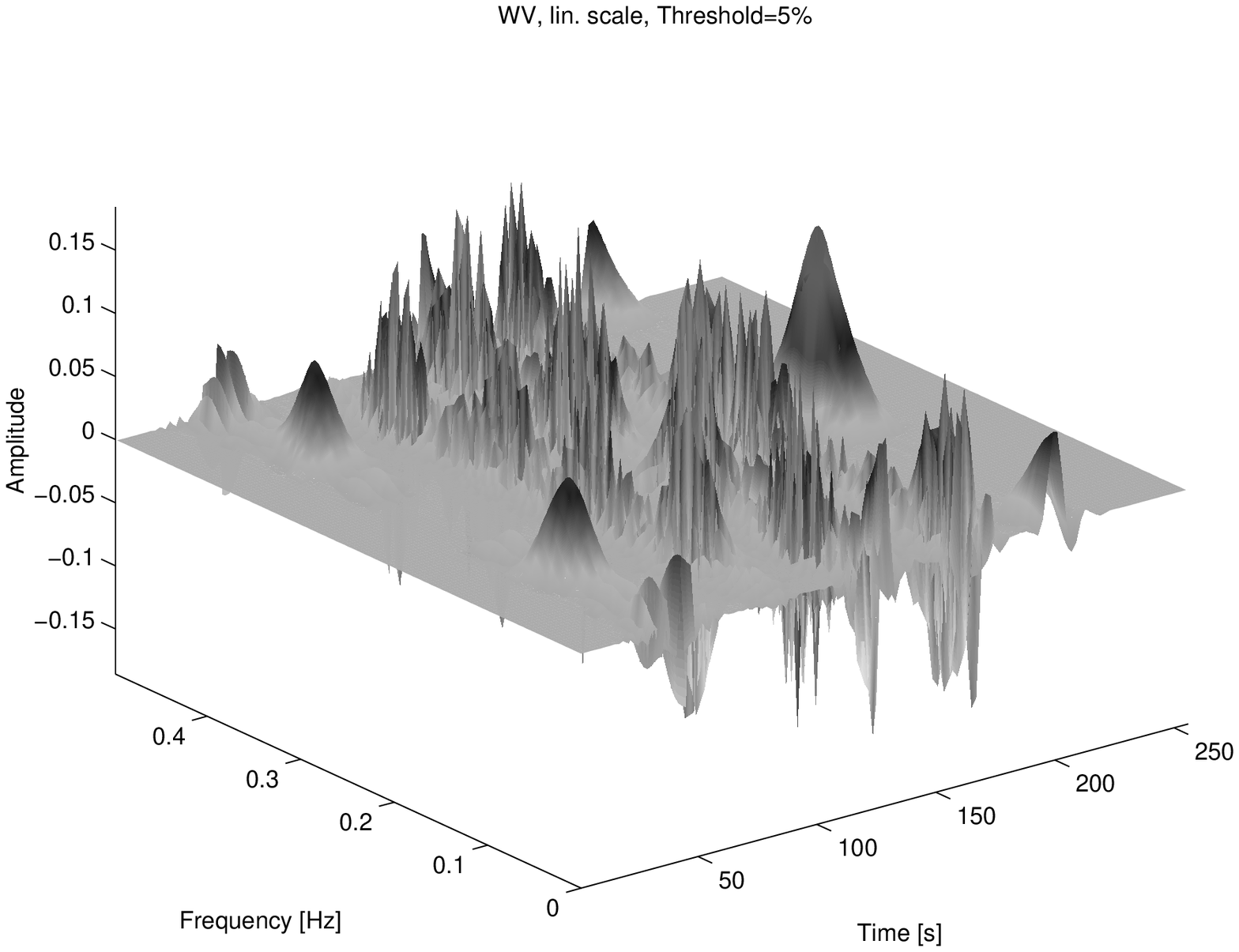}
\end{tabular}
\end{center}
\caption{Wigner function for 3 wavelet packets.}
\end{figure}

\newpage
\twocolumn

\begin{figure}
\begin{center}
\begin{tabular}{c}                                                                                       
\includegraphics*[width=45mm]{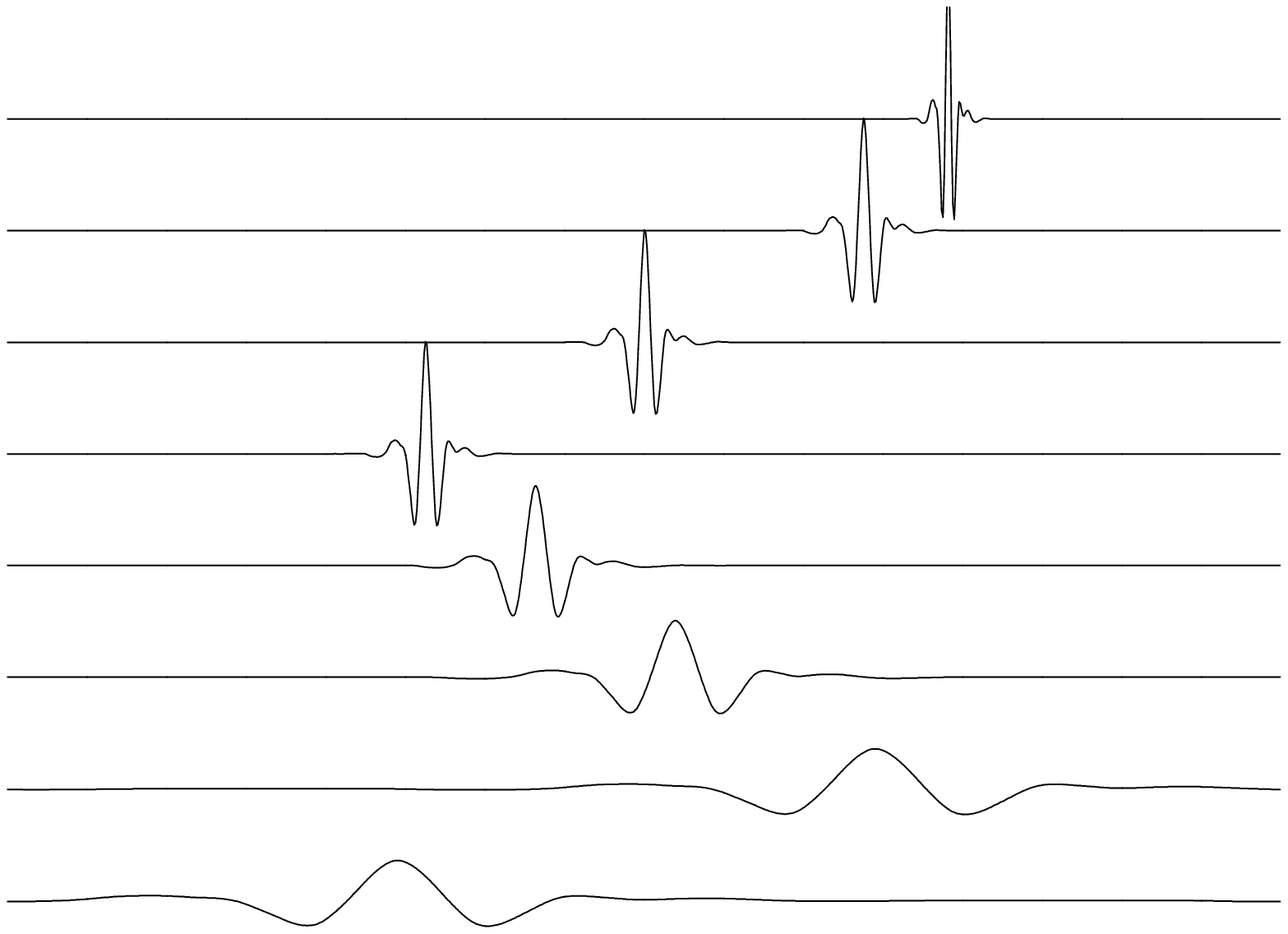}
\end{tabular}
\end{center}                                                          
\caption{1-dimensional MRA modes.}                                                                       
\end{figure}

\begin{figure}
\begin{center}
\begin{tabular}{c}                                                                                         
\includegraphics*[width=60mm]{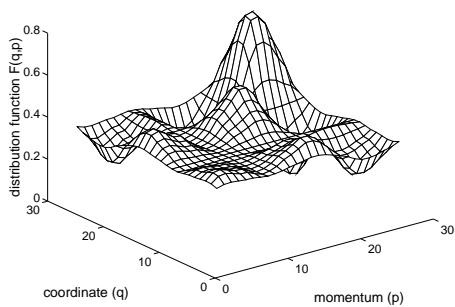} 
\end{tabular}
\end{center}                                   
\caption{2-dimensional MRA mode.}                                                                                 
\end{figure}

\begin{figure}
\begin{center}
\begin{tabular}{c}  
\includegraphics*[width=60mm]{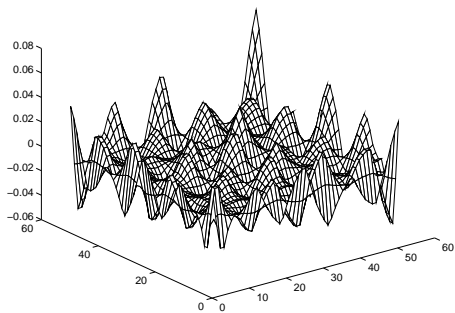}
\end{tabular}
\end{center}
\caption{Contribution to WF at scale 1.}
\end{figure}

\begin{figure}                                                                                               
\begin{center}
\begin{tabular}{c}                                                                                        
\includegraphics*[width=60mm]{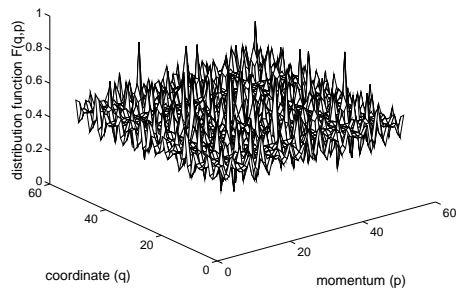}
\end{tabular}
\end{center}                                                          
\caption{Chaotic-like pattern.}                                                                  
\end{figure}

\newpage

\begin{figure}
\begin{center}
\begin{tabular}{c}
\includegraphics*[width=60mm]{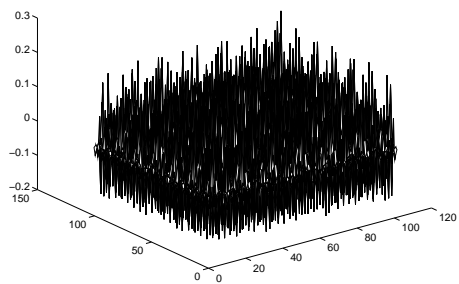}
\end{tabular}
\end{center}
\caption{Entangled state/pattern.}
\end{figure}

\begin{figure}                                                                             
 \begin{center}
\begin{tabular}{c}                                                                                  
\includegraphics*[width=60mm]{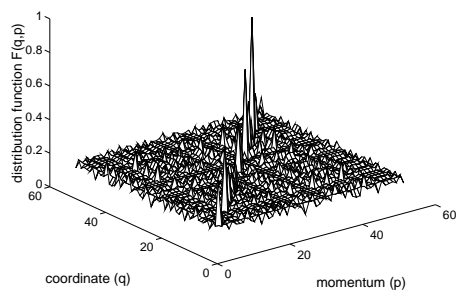} 
\end{tabular}
\end{center}                                                         
\caption{Localized pattern: waveleton.}                                                          
\end{figure}

\end{document}